\documentclass{emulateapj}

\usepackage{graphicx} 
\usepackage{amsmath}
\usepackage{hyperref}

\begin{document}

\title{First Detection of Thermal Radio Emission from Solar-Type Stars\\with the Karl G. Jansky Very Large Array}
\shorttitle{Thermal Radio Emission from Solar-Type Stars}

\shortauthors{Villadsen et al.}
\author{Jackie Villadsen\altaffilmark{1,2}, Gregg Hallinan\altaffilmark{2}, Stephen Bourke\altaffilmark{2}, Manuel G\"udel\altaffilmark{3}, Michael Rupen\altaffilmark{4}}
\altaffiltext{1}{Corresponding author: jrv@astro.caltech.edu}
\altaffiltext{2}{Department of Astronomy, California Institute of Technology, 1200 E. California Ave., Pasadena, CA 91125, USA}
\altaffiltext{3}{Department of Astrophysics, University of Vienna, T\"urkenschanzstrasse 17, A-1180 Vienna, Austria}
\altaffiltext{4}{National Radio Astronomy Observatory, Socorro, NM 87801, USA}

\begin{abstract}
We present the first detections of thermal radio emission from the atmospheres of solar-type stars $\tau$ Cet, $\eta$ Cas A, and 40 Eri A. These stars all resemble the Sun in age and level of magnetic activity, as indicated by X-ray luminosity and chromospheric emission in calcium-II H and K lines. We observed these stars with the Karl G. Jansky Very Large Array with sensitivities of a few $\mu$Jy at combinations of 10.0, 15.0, and 34.5\,GHz. $\tau$ Cet, $\eta$ Cas A, and 40 Eri A are all detected at 34.5\,GHz with signal-to-noise ratios of 6.5, 5.2, and 4.5, respectively.  15.0-GHz upper limits imply a rising spectral index greater than 1.0 for $\tau$ Cet and 1.6 for $\eta$ Cas A, at the 95\% confidence level. The measured 34.5-GHz flux densities correspond to stellar disk-averaged brightness temperatures of roughly 10,000\,K, similar to the solar brightness temperature at the same frequency. We explain this emission as optically-thick thermal free-free emission from the chromosphere, with possible contributions from coronal gyroresonance emission above active regions and coronal free-free emission.  These and similar quality data on other nearby solar-type stars, when combined with ALMA observations, will enable the construction of temperature profiles of their chromospheres and lower transition regions.
\end{abstract}



\section{Introduction}

Efforts to measure radio emission from the Sun started as early as the 1890s by \cite{wilsing1896}.  The first detections occurred during the boom in radio technology development during World War II, when military radio engineers James Stanley \cite{hey1946} in Britain, George Clark \cite{southworth1945} in the United States, and Bruce Slee in Australia \citep{orchiston2005} all independently identified solar radio emission as a source of interference in their radar signals.

Seventy years later, solar radio observations have contributed significantly to a detailed (although far from complete) understanding of the solar atmosphere.  Solar flares produce transient radio emission from MHz to GHz frequencies, including gyrosynchrotron storms and coherent bursts, which act as diagnostics of electron density and magnetic field strength in the solar corona (see \cite{bastian1998} for a review).

From centimeter to far-infrared wavelengths, the quiet Sun emits optically-thick thermal radiation that departs slightly from the Rayleigh-Jeans law due to variation of brightness temperature with frequency.  In models of the solar atmosphere (e.g., \cite{loukitcheva2004}), the quiet-Sun brightness temperature spectrum probes atmospheric temperature from the temperature minimum (far-IR) to the upper chromosphere at $\sim10^4$\,K (cm wavelengths), where free-free absorption dominates the opacity.  During periods of heightened solar activity, low-frequency solar radiation (below 10-20\,GHz) is enhanced by bright spots above active regions, where the $\sim10^6$-K corona is optically thick due to gyroresonant and/or free-free opacity.  For this reason, the 10.7-cm solar flux density $F_{10.7}$ is a traditional measure of solar activity, varying by a factor of 2 to 3 during the solar cycle.

For the nearest stars, radio luminosities at the level of the quiet Sun correspond to $\mu$Jy flux densities at GHz frequencies, rising to tens of $\mu$Jy above 20\,GHz.  Previous generations of radio telescopes have not been sensitive enough to detect such a signal.  Probing higher luminosities, \cite{drake1993} detected 8.3-GHz thermal chromospheric emission from Procyon, a slightly-evolved F5 subgiant, which has a radio luminosity roughly 10 times that of the quiet Sun due to its larger surface area and elevated brightness temperature.

\setlength{\tabcolsep}{0.05in}
\begin{deluxetable*}{llllllllllllll}
\tablecaption{Basic stellar properties. \label{table:basic}}
\tablehead{
	\colhead{Name} &
	\colhead{HD} &
	\colhead{Dist.} &
	\colhead{Spectral} &
	\colhead{Ref.} &
	\colhead{Mass} &
	\colhead{Ref.} &
	\colhead{Radius} &
	\colhead{Ref.} &
	\colhead{$\mathrm{T_{eff}}$} &
	\colhead{Ref.} &
	\colhead{[Fe/H]} &
	\colhead{Ref.} &
	\colhead{Known Stellar Companions} \\
	\colhead{} & 	
	\colhead{} & 	
	\colhead{(pc)} & 
	\colhead{Type} & 	
	\colhead{} &	
	\colhead{(M$_\odot$)} & 
	\colhead{} &
	\colhead{(R$_\odot$)} &
	\colhead{} &
	\colhead{(K)} &
	\colhead{} &
	\colhead{} &
	\colhead{} &
	\colhead{}
}
\startdata
$\tau$ Cet	& 10700	& 3.65	& G8.5V	& 1	& 0.783	& 3	& 0.790	& 6	& 5400	& 1	& -0.40	& 1	& none \\
$\eta$ Cas A	& 4614	& 5.95	& F9V	& 2	& 0.972	& 4	& 1.039	& 4	& 6000	& 4	& -0.25 & 8	& K7V @ 70 AU (12")~\footnote{\cite{usno}.} \\
40 Eri A	& 26965	& 4.98 & K0.5V	& 1	& 0.84	& 5	& 0.77	& 7	& 5100	& 1	& -0.28	& 1	& DA3 \& M5Ve @ 400 AU (80")~\footnote{\cite{heintz1974}.} \\ [1.5ex]
Sun & \nodata & 4.8e-6 & G2V & & 1 & & 1 & & 5700 & & 0 & & none
\enddata
\tablecomments{
Distances to stars are based on parallaxes from \cite{vanLeeuwen2007}.
The orbital separations of any stellar companions are reported as the orbital semi-major axis. 
References: 
(1) \cite{gray2006},
(2) \cite{gray2001},
(3) \cite{teixeira2009},
(4) \cite{boyajian2012},
(5) \cite{holmberg2007},
(6) \cite{diFolco2007},
(7) \cite{demory2009},
(8) \cite{gray2001b}.
}
\end{deluxetable*}

To date, all radio-detected main-sequence stars have microwave luminosities orders of magnitude higher than the quiet Sun (for a review of stellar radio emission, refer to \cite{gudel2002}). \cite{gary1981} and \cite{linsky1983} reported and interpreted the first detections of radio emission from low-mass main-sequence stars.   The stars' high radio luminosities are predominantly attributed to gyrosynchrotron emission from a persistent non-thermal electron population in the corona, a feature with no analog in the non-flaring Sun.  This intense radio emission is accompanied by vigorous magnetic activity, indicated by X-ray luminosities orders of magnitude above solar.  \cite{gudel&benz1993} observed a correlation between X-ray and radio luminosity in quiescent emission from coronae of magnetically-active dwarf stars.  \cite{benz&gudel1994} observed that the X-ray-radio luminosity relation is also seen in solar flares, suggesting that non-thermal stellar radio ``coronae'' may consist of the emission from many small flares, or at least are continuously heated by flares.  \cite{gudel1994} detected 8.5-GHz emission from X-ray-bright solar-type stars; these stars have radio luminosities a few thousand times that of the Sun, consistent with the X-ray-radio luminosity relation, so their radio luminosity is attributed to gyrosynchrotron emission.

The X-ray-radio luminosity relationship observed in active stars does not extend to stars with moderate-to-low magnetic activity.  \cite{gudel1998} and \cite{gaidos2000} used the Very Large Array to search for 8.4-GHz radio emission from young, moderately active solar-type stars $\pi^1$ UMa, $\kappa^1$ Cet, and $\beta$ Com at 8.4\,GHz, reaching 3$\sigma$ detection limits of 20-30\,$\mu$Jy.  These upper limits correspond to radio luminosities of $\sim10^{12.5}$ erg/s, which fall below expectations based on their X-ray luminosities.  These stars do not show evidence of the strong gyrosynchrotron component that outshines thermal emission in more active stars, but their thermal component is too faint to detect due to their distance (9 to 14 pc).

To detect the analog of the thermal radiation that dominates quiescent microwave emission from the Sun, we turn to nearby solar-type stars (at 4 to 6 pc) with solar-like X-ray luminosities.  Previous to the results reported here, the most sensitive microwave observations of such stars were performed with the Very Large Array by \cite{gudel1992}, placing a 3$\sigma$ upper limit of 80\,$\mu$Jy on the 8.3-GHz flux density of 40 Eridani A. (In contrast, at sub-mm wavelengths, the optically-thick thermal emission from the low chromospheres of these stars is of order a few mJy, but is blended with debris-disk emission of comparable brightness, as in the case of the observations of $\tau$ Cet's debris disk by \cite{greaves2004}.)  With the enhanced sensitivity of the Karl G. Jansky Very Large Array (VLA), the observations described in this paper reached a 3$\sigma$ detection limit of 6 to 12\,$\mu$Jy in a few hours of observation, enabling the first detections of microwave radio emission from the thermal atmospheres of solar-type stars.  (The term ``solar-type stars,'' as used in this paper, refers to main sequence stars of spectral type late F through early K.)

\setlength{\tabcolsep}{0.04in}
\begin{deluxetable}{lllllllll}
\tablecaption{Measures of stellar magnetic activity.\label{table:activity}}
\tablehead{
	\colhead{Name} &
	\colhead{$\log_{10} \mathrm{L_X}$} &
	\colhead{Ref.} &
	\colhead{$\mathrm{P_{rot}}$} &
	\colhead{Ref.} &
	\colhead{$\log_{10} \mathrm{R'_{HK}}$} &
	\colhead{Ref.} &
	\colhead{Age} &
	\colhead{Ref.} \\
	\colhead{} &
	\colhead{(erg~s$^{-1}$)} &
	\colhead{} &
	\colhead{(days)} &
	\colhead{} &
	\colhead{} &
	\colhead{} &
	\colhead{(Gyr)} &
	\colhead{}
}
\startdata
$\tau$ Cet	& 26.5	& 1	& 34	& 3	& -5.01	 & 6	& 5.8	& 7 \\
$\eta$ Cas A	& 27.4	& 1	& 17	& 4	& -4.93	 & 6	& 2.9	& 7 \\
40 Eri A	& 27.2	& 1	& 43	& 3	& -4.872 & 3	& 5.6	& 7 \\ [1.5ex]
Sun 		& 27.35	& 2	& 26.1	& 5	& -4.906 & 7	& 4.6	& 8
\enddata
\tablecomments{
References: 
(1) NEXXUS catalog \citep{nexxus}. Stellar X-ray luminosities from pointed observations by the ROSAT High-Resolution Imager (HRI).
(2) \cite{judge2003}. Solar $\mathrm{L_X}$ averaged over the solar cycle, calculated for the ROSAT All-Sky Survey 0.1 to 2.4 keV bandpass. The solar $\mathrm{L_X}$ varies by about an order of magnitude from solar minimum to solar maximum.
(3) \cite{baliunas1996}. Rotation periods determined from rotational modulation of Ca II H\&K lines.
(4) \cite{wright2004}.  Rotation period inferred from $\log_{10} \mathrm{R'_{HK}}$ by activity-rotation relation.
(5) \cite{donahue1996}.
(6) \cite{cantomartins2011}.
(7) \cite{mamajek2008}. Note that age is not an independent measure of activity: these ages were derived using empirical gyrochronology relations from rotation rates, which were in turn derived from $\log_{10} \mathrm{R'_{HK}}$.
(8) \cite{bouvier2010}.
}
\end{deluxetable}

In this paper we present the first detections of thermal radio emission from three nearby solar-type stars with solar-like levels of magnetic activity: $\tau$ Ceti, $\eta$ Cassiopeiae A, and 40 Eridani A.  In Section \ref{section:obs}, we review our sample and describe the observing program.  Section \ref{section:results} presents the source detections and upper limits and compares them to the solar brightness temperature spectrum. Section \ref{section:interpret} discusses a variety of possible emission mechanisms for the detected radiation.  In Section \ref{section:conclusions}, we conclude with a review of the detected sources and most likely emission mechanisms and discuss the potential for future observations with ALMA and the SKA.

\section{Observations}\label{section:obs}

Our sample consists of three of the nearest stars of spectral type F9V through K0.5V that are observable from the VLA's latitude: $\tau$ Cet, $\eta$ Cas A, and 40 Eri A. Tables \ref{table:basic} and \ref{table:activity} compare these stars' properties, including a number of measures of stellar activity, to those of the Sun. All three stars are a good match for the Sun in age and activity level.

\setlength{\tabcolsep}{0.06in}
\begin{deluxetable*}{lccccclccll}
\tablecaption{Summary of observations.\label{table:obs}}
\tablehead{
	\colhead{} &
	\colhead{} &
	\colhead{Center} &
	\colhead{Band-} &
	\colhead{\# of} &
	\colhead{Time on} &
	\colhead{VLA} &
	\colhead{Synthesized Beam} &
	\colhead{} &
	\colhead{Phase} &
	\colhead{Flux} \\
	\colhead{Star} &
	\colhead{Band} &
	\colhead{Frequency} & 
	\colhead{width} &
	\colhead{Epochs} &
	\colhead{Source} & 
	\colhead{Configuration} &
	\colhead{Dimensions} &
	\colhead{RMS} & 
	\colhead{Calibrator} &
	\colhead{Calibrator} \\
	\colhead{} &
	\colhead{} &
	\colhead{(GHz)} & 
	\colhead{(GHz)} &
	\colhead{} &
	\colhead{(h)} & 
	\colhead{} &
	\colhead{(FWHM)} &
	\colhead{($\mathrm{\mu}$Jy)} & 
	\colhead{} &
	\colhead{} 
}
\startdata
$\tau$ Cet & Ka & 34.5 & 8.0 & 2 & 5.0 & DnC & 2.11'' $\times$ 1.45'' & 3.9 & J0204-1701 & 3C147 \\
 & Ku & 15.0 & 6.0 & 1 & 2.0 & D & 8.99'' $\times$ 4.64'' & 3.0 \\ [1.5ex]
$\eta$ Cas A & Ka & 34.5 & 8.0 & 4 & 8.0 & DnC/C & 1.01'' $\times$ 0.83'' & 3.1 & J0102+5824 & 3C147 \\ 
 & Ku & 15.0 & 6.0 & 3 & 5.0 & D/C & 2.40'' $\times$ 2.25'' & 2.1 & & \& 3C48 \\
 & X & 10.0 & 4.0 & 2 & 3.0 & D & 11.93'' $\times$ 7.98'' & 2.7 \\ [1.5ex]
40 Eri A & Ka & 34.5 & 8.0 & 3 & 5.5 & DnC$\rightarrow$C/C & 0.98'' $\times$ 0.73'' & 3.7 & J0423-0120 & 3C147
\enddata
\tablecomments{All observations were made between March and September 2013.}
\end{deluxetable*}

Table \ref{table:obs} summarizes the observations of all stars in the sample. Each star in the sample was observed with the full VLA array of 27 antennas in X band (8.0\,-\,12.0\,GHz), Ku band (12.0\,-\,18.0\,GHz), and/or Ka band (tuned to 30.5\,-\,38.5\,GHz). Observations of a single source in different bands occurred on different dates.  The VLA WIDAR correlator's 3-bit observing mode enabled up to 8-GHz bandwidth. Observations were performed between March and September 2013, with the VLA in D configuration and C configuration as well as intermediate configurations.  Observations alternated between a nearby phase calibrator and the target source with cycle times of 7.5 minutes in Ka band and 17 minutes in X and Ku band.  Typical sensitivities obtained with the full bandwidth in one hour on source were 4\,$\mu$Jy rms in X and Ku bands and 7\,$\mu$Jy rms in Ka band.

Typical VLA observations in X, Ku, and Ka bands reach an absolute flux calibration accuracy of 5\% to 10\%.  Throughout the rest of the paper, quoted measurement errors reflect only the statistical error on measurements due to random noise on the source visibilities, not the systematic error due to absolute flux calibration.  Measured flux density may also be reduced compared to the true value if the gain calibrations interpolated from the phase calibrator are not sufficient to correct for the variation of gain phases with time; however, the gain solutions obtained for the phase calibrator in all bands varied slowly and smoothly over time, suggesting that this source of error is negligible.

\subsection{Source Motion\label{section:motion}}
The expected positions of the sources were determined using Hipparcos coordinates and proper motions \citep{vanLeeuwen2007}, with an additional correction for parallactic motion based on the distances in Table \ref{table:basic}.  In addition, the position of $\eta$ Cas A was corrected for displacement due to orbital acceleration, which resulted in a 0.3'' displacement northwest of the position expected from proper motion and parallax. The orbital ephemeris calculation used an Excel workbook developed by Brian Workman\footnote{As of 2014 January 15, available online at \url{http://www.saguaroastro.org/content/db/binaries_6th_Excel97.zip}.} with data from the US Naval Observatory's Washington Double Star Catalog by \cite{usno}.  40 Eri A did not require a correction for orbital motion because the 40 Eri orbital period is much longer than that of $\eta$ Cas.

Since the targets are within a few parsecs, the sources moved by as much as 1'' due to proper motion and parallax during the half-year in which we observed.  In comparison, our Ka-band observations reached a typical synthesized beam of 1'' and an astrometric accuracy of order 0.1''.  To avoid smearing of the source and obtain accurate astrometry, we used the interferometry software package CASA's routine \textit{fixvis} to shift the visibility phases to keep the expected location of the source at the phase center in all observations, before combining and imaging visibility data from different epochs.

\subsection{Chance Alignment of Extragalactic Sources\label{section:confusion}}
Based on the 3-GHz source counts of \cite{condon2012}, chance alignment of an unrelated source is a negligible source of error in these observations. \cite{condon2012} performed 3-GHz source counts with 1-$\mu$Jy sensitivity and measured a differential source count of:
\begin{equation}
n(S) = 9000 S^{-1.7} \mathrm{~Jy^{-1}~sr^{-1}}.
\end{equation}

Considering the scenario where source counts are independent of frequency, then the highest probability of finding a 1$\sigma$ unrelated source within the synthesized beam is 18\%, for the X-band observations of $\eta$ Cas A (since those observations have the largest synthesized beam). For most other observations, the probability is much lower. In all cases where a source was detected, the probability of an unrelated source with flux density greater than or equal to the detected level (see Table \ref{table:detections}) falling in the synthesized beam is 0.1\% or less.

We base our statistics on 3-GHz source counts because source counts at higher frequencies are not available for $\mu$Jy-level sources, but the probabilities of chance alignment of detectable sources in the observed bands (10, 15, and 34.5\,GHz) are likely lower than at 3\,GHz.  The Australia Telescope 20\,GHz Survey, which surveyed the 20-GHz southern sky down to a depth of 40\,mJy, found that only 1.2\% of 20-GHz sources were undetected or weakly detected at 5\,GHz \citep{murphy2010} and that the majority of 5- and 8-GHz sources are flat- or falling-spectrum \citep{massardi2011}.  The 15.7-GHz Tenth Cambridge survey with the Arcminute Microkelvin Imager \citep{whittam2013}, which was complete to 500\,$\mu$Jy over 12\,deg$^2$, found that sources below 800\,$\mu$Jy tend to be flat spectrum.  Thus it is reasonable to use 3-GHz source counts to put upper limits on the probability of chance alignment of extragalactic sources in our observations.

\section{Results}\label{section:results}
\subsection{Detections \label{section:detections}}

\setlength{\tabcolsep}{0.06in}
\begin{deluxetable}{lccccr}
\tablecaption{Detections.\label{table:detections}}
\tablehead{
	\colhead{} &
	\colhead{34.5-GHz} &
	\colhead{} &
	\colhead{} &
	\colhead{Position} &
	\colhead{Brightness} \\
	\colhead{Star} &
	\colhead{Flux Density} &
	\colhead{RMS} &
	\colhead{SNR} &
	\colhead{Offset\,\tablenotemark{a}} &
	\colhead{Temperature\,\tablenotemark{b}} \\
	\colhead{} &
	\colhead{($\mu$Jy)} &
	\colhead{($\mu$Jy)} &
	\colhead{} &
	\colhead{(sigma)} &
	\colhead{(K)}
}
\startdata
$\tau$ Cet & 25.3 & 3.9 & 6.5 & 3.9 & 9300$\pm$1400\\
$\eta$ Cas A & 16.0 & 3.1 & 5.2 & 0.4 & 9500$\pm$1800\\
40 Eri A & 16.5 & 3.7 & 4.5 & 0.5 & 10200$\pm$2300
\enddata
\tablenotetext{a}{The offset from the expected position based on Hipparcos astrometry \citep{vanLeeuwen2007}, in units of sigma (i.e. the distance between the measured and expected coords, divided by the amplitude of the error ellipse in the offset direction).}
\tablenotetext{b}{Brightness temperature is averaged over the stellar disk, using the photospheric radii and distances reported in Table \ref{table:basic}.  The reported errors do not include the 5\%- to 10\%-percent errors typical of absolute flux calibration with the VLA for X, Ku, and Ka bands.}
\end{deluxetable}

\setlength{\tabcolsep}{0.06in}
\begin{deluxetable}{lcccc}
\tablecaption{Non-detections.\label{table:ULs}}
\tablehead{
	\colhead{} &
	\colhead{Center} &
	\colhead{} &
	\colhead{95\%-Conf.} &
	\colhead{99\%-Conf.} \\
	\colhead{Star} &
	\colhead{Frequency} &
	\colhead{RMS} &
	\colhead{Upper Limit} &
	\colhead{Upper Limit} \\
	\colhead{} &
	\colhead{(GHz)} &
	\colhead{($\mu$Jy)} &
	\colhead{($\mu$Jy)} &
	\colhead{($\mu$Jy)}
}
\startdata
$\tau$ Cet & 15.0 & 3.0 & 9.6 & 11.7 \\
$\eta$ Cas A & 10.0 & 2.7 & 6.6 & 8.4 \\
$\eta$ Cas A & 15.0 & 2.1 & 3.8 & 5.0
\enddata
\end{deluxetable}

\begin{figure}
  \begin{center}
     \includegraphics{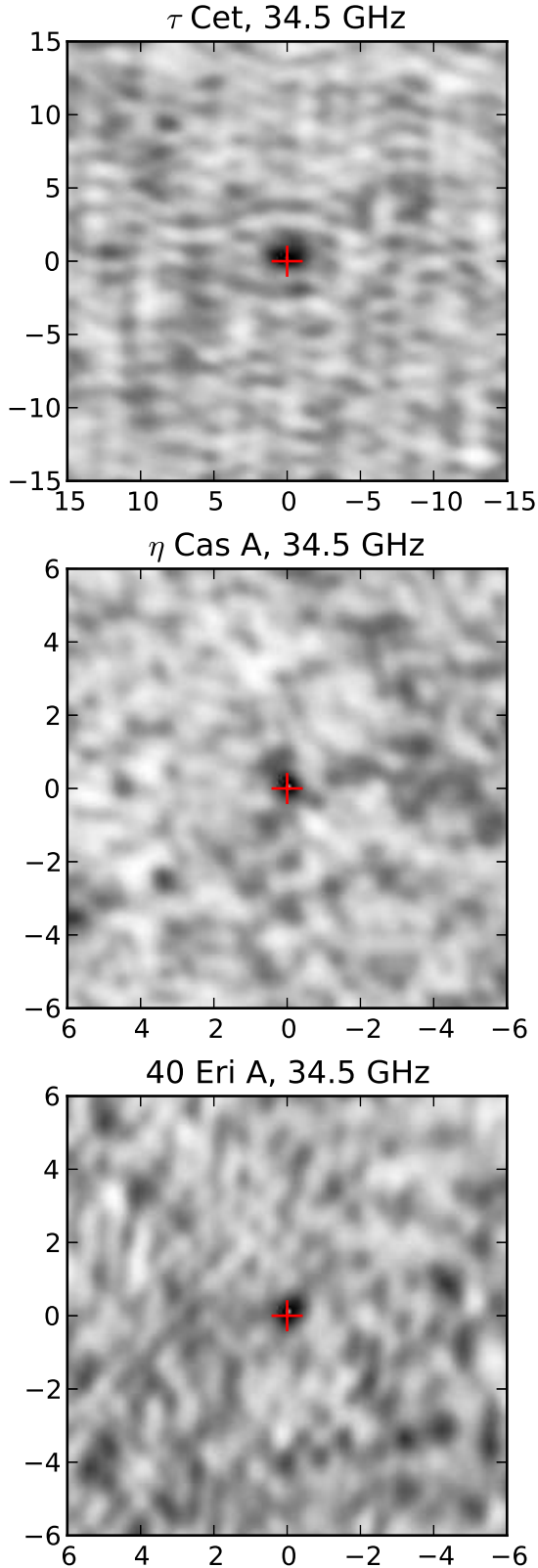}
  \end{center}
  \caption{Ka-band images of the observed stars.  The x- and y-axis are labelled with offset in arcseconds in the E-W and N-S directions, respectively.  The cross (colored red in the online version) shows the location of the star, determined from Hipparcos astrometry~\citep{vanLeeuwen2007} and adjusted to the epoch of the observations accounting for proper motion, parallax, and (for $\eta$ Cas A) orbital motion.  The size of the cross marker is arbitrary because the errors on astrometry, of order 0.1'', are too small to show in these images. See Section \ref{sec:positions} for a comparison of Hipparcos positions and observed positions. \label{fig:detections}}
\end{figure}

Solar-type stars $\tau$ Cet, $\eta$ Cas A, and 40 Eri A were all detected in Ka band (center frequency 34.5\,GHz) with signal-to-noise ratio (SNR) of 6.5, 5.2, and 4.5, respectively.  Table \ref{table:detections} gives the measured flux density for each source, where all the Ka-band observations of a single source from different dates have been combined. Figure \ref{fig:detections} shows the VLA images of the detections (also one image per star, with all Ka-band observations combined).  Flux densities and source positions were determined by using CASA's \textit{uvmodelfit} task to fit a point source model to the visibilities.  To check whether the source was consistent with a point source, CASA's \textit{imfit} task was also used to fit an elliptical Gaussian with unconstrained shape, which yielded dimensions consistent with the CLEAN beam to within 3 sigma for all sources.

The sources were imaged in full Stokes, but were only detected at or above the 3$\sigma$ level in Stokes I (total intensity).  We used the Stokes V (circularly polarized) flux density at the pixel representing the expected position of the source to calculate confidence intervals on the degree of circular polarization $r_c=V/I$ for each source, where $r_c$ varies from -1 to 1.  We obtained 95\%-confidence intervals on $r_c$ of [-0.52,0.14], [-0.18,0.69], and [-0.87,0.12] for $\tau$ Cet, $\eta$ Cas A, and 40 Eri A, respectively.

To check for flares, we created Ka-band time series in Stokes I and V for each of the stars, averaging the observations over a range of timescales from 15 seconds to 25 minutes.  Analysis of the time series showed no statistically significant evidence of variability, consistent with quiescence.  All time series were consistent with constant flux density with reduced $\chi^2$ ranging from 1 to 2 for different averaging times, with values of 1.1, 1.3, and 1.6 for 25-minute averaging of Stokes I for $\tau$ Cet, $\eta$ Cas A, and 40 Eri A, respectively.  Thus, these detections are consistent with non-flaring emission.

\subsubsection{Positions\label{sec:positions}}

The fifth column of Table \ref{table:detections}, titled ``Position Offset,'' gives the offset of the VLA-observed position of each source compared to the phase center, where the phase center is the expected position predicted from Hipparcos coordinates and proper motions, as described in Section \ref{section:motion}.  This offset is given in units of sigma, i.e. the offset distance in arcseconds divided by the amplitude of the position error ellipse in arcseconds in the offset direction.  The position error ellipse is the synthesized beam, defined as the CLEAN restoring beam, divided by the signal-to-noise ratio of the source detection.  $\eta$ Cas A and 40 Eri A show good agreement between the expected and observed positions.  $\tau$ Cet shows a 3.9$\sigma$, or 0.37'', difference; however, the expected position agrees well with the location of peak flux density in the image.  We are confident that the observed source is indeed $\tau$ Cet, because of the low probability of coincidence of an extragalactic source, and because the observed source is the only source detected in the 1.3' primary beam at the 4$\sigma$ level or greater.

\subsection{Upper Limits}
Our observations at 10 GHz ($\eta$ Cas A) and 15 GHz ($\tau$ Cet and $\eta$ Cas A) did not lead to detections, but enabled us to put upper limits on the flux density from these sources at these frequencies.  In the case of a non-detection, the flux density in the pixel at the star's known location still constitutes a measurement of the flux density, where the measurement is drawn from a Gaussian distribution whose mean is the true flux density and whose standard deviation is the image RMS.  The measured flux densities were 4.6\,$\mu$Jy for $\tau$ Cet in Ku band, -0.56\,$\mu$Jy for $\eta$ Cas A in Ku band, and 1.8\,$\mu$Jy for $\eta$ Cas A in X band.  The posterior probability distributions for flux density were calculated using flat priors that disallowed negative values for the true flux density of the source, resulting in the 95\%- and 99\%-confidence upper limits on flux density shown in Table~\ref{table:ULs}.

For comparison, the traditional method of reporting 3$\sigma$ upper limits is equivalent to giving the 99.73\%-confidence upper limit in the case where the flux density measured at the known location of the source is zero (or where the location of the source is not known).

\subsection{Brightness Temperature Spectra}\label{section:SEDs}

Figure \ref{fig:SEDs} shows our constraints on the flux density and brightness temperature ($T_b$) spectra of the observed stars, compared to the solar spectrum.  All three detections of solar-type stars are consistent with the solar brightness temperature in Ka band, suggesting that the emission observed in these cases is most likely chromospheric blackbody emission as in the Sun (see Section \ref{section:chromo}).  The upper limits placed on the stars' $T_b$ are consistent with the solar brightness temperature spectrum.

\subsubsection{Spectral Index}
\begin{figure*}
  \begin{center}
     \includegraphics{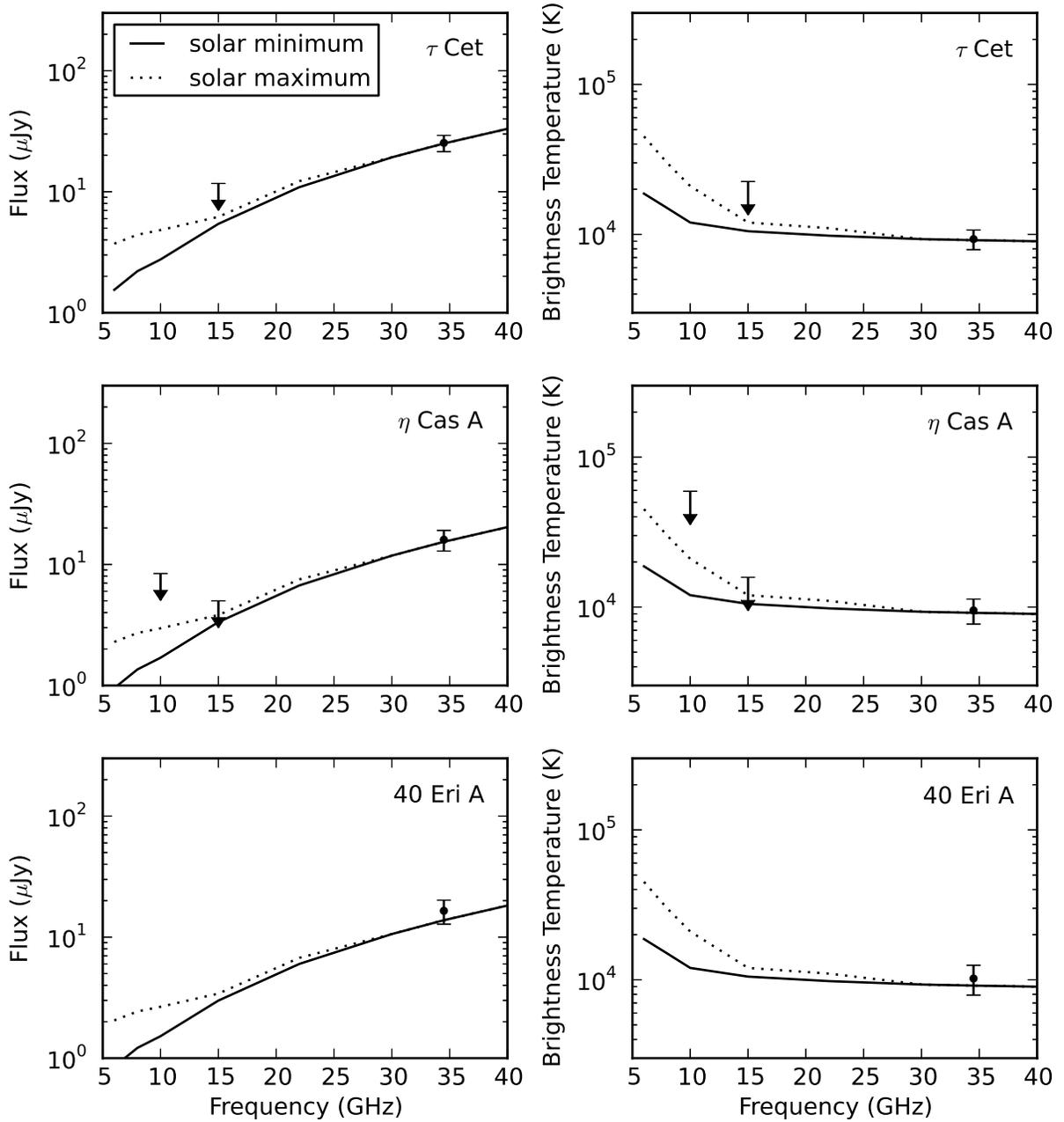}
  \end{center}
   \caption{Left column: Flux spectra for the observed stars. Right column: Stellar disk-averaged brightness temperature spectra.  Both columns: non-detections are shown as downwards arrows marking the 99\%-confidence upper limits (Table~\ref{table:ULs}) and detections as points with 1$\sigma$ error bars (Table~\ref{table:detections}).  The plots also show the solar spectra from \cite{white2004} for solar minimum (solid line) and solar maximum (dotted line); in the left plots the solar flux density is scaled to the distance and radius of each star, so the lines show the flux density the star would have if it had solar brightness temperatures. The solar cycle mainly affects the solar spectrum below 15\,GHz, where gyroresonant emission from active regions contributes significantly to the emission, because the number and strength of active regions varies with the solar cycle. Above 15\,GHz, solar radiation is predominantly chromospheric thermal free-free emission, which is constant throughout the solar cycle because the temperature of the chromosphere remains steady. \label{fig:SEDs}}
\end{figure*}

The combination of the upper limits on 15.0-GHz flux density and the measured 34.5-GHz flux densities indicates that $\tau$ Cet and $\eta$ Cas A have rising spectra from 15.0 to 34.5\,GHz (as seen on the left in Figure~\ref{fig:SEDs}).  We calculated lower limits on the 15-to-34.5-GHz spectral index $\alpha$ for these two stars, where $\alpha$ is defined as the positive power-law index for the flux spectrum:
\begin{equation}
S_\nu \propto \nu^\alpha .
\end{equation}
To do so, we considered the measured value for the Ku-band flux density to be the Ku-band flux density in the pixel at the star's known location (4.6\,$\mu$Jy for $\tau$ Cet and -0.56\,$\mu$Jy for $\eta$ Cas A), drawn from a Gaussian distribution with the image RMS.  The posterior probability distribution for spectral index was calculated using least-informative Jeffreys priors, resulting in 95\%-confidence lower limits on $\alpha$ of 1.0 and 1.6, for $\tau$ Cet and $\eta$ Cas A, respectively, and 99\%-confidence lower limits of 0.8 and 1.1.

It should be noted that these statistical constraints on spectral index reflect only the random error on the visibilities: as noted in Section~\ref{section:obs}, we do not account for the systematic effect of relative errors in absolute flux calibration for different bands.

\section{Discussion}\label{section:interpret}

\subsection{Chromospheric Stellar Disk Emission}\label{section:chromo}
35-GHz solar radio emission is dominated by thermal emission from the chromosphere, which is optically thick at cm wavelengths due to free-free opacity.  The disk-averaged brightness temperatures reported in Table \ref{table:detections} are consistent with the 35-GHz quiet-Sun brightness temperature of 9300\,K reported in \cite{white2004}.

At microwave frequencies, the radio opacity of the quiet Sun is dominated by free-free opacity from free electrons and ions.  At an upper-chromosphere temperature of $10^4$\,K and frequencies near 34.5 GHz, the free-free absorption coefficient (from Equation 20 in \cite{dulk1985}) is:
\begin{equation}
\kappa_{\nu,ff} \approx 0.076 \frac{n_e^2}{\nu^2 T^{3/2}},
\label{eqn:abs_coeff}
\end{equation}
where all quantities are in cgs units.
The frequency dependence of free-free opacity implies that different frequencies probe different depths, and therefore different temperatures, in the stellar atmosphere.  Since free-free opacity also depends on density, a brightness temperature spectrum can constrain temperature and density profiles of model chromospheres, as for models of the solar chromosphere in \cite{loukitcheva2004} and \cite{fontenla2007}.  Comparable models of the atmospheres of these low-activity solar-type stars may be constrained using a brightness temperature spectrum consisting of the VLA data presented here combined with ALMA and far-IR data.  The cm-wavelength observations in this paper probe the upper chromosphere and the base of the transition region, whereas mm-wavelength data would provide a glimpse of the lower chromosphere, and far-IR data would reveal the conditions of the temperature minimum (see for example the direct detection of the $\alpha$ Cen A temperature minimum by \cite{liseau2013}).

\cite{lim1998} took a similar approach to construct a temperature profile of the atmosphere of supergiant Betelgeuse, observing with the VLA from 5 to 43\,GHz.  The supergiant star's extended atmosphere was resolved, enabling direct measurement of the atmospheric height probed by each frequency and of the source size for the purpose of calculating brightness temperature. In our case, the source is unresolved, but the chromosphere is likely close enough to the photosphere that assuming the photospheric radius will yield an accurate calculation of brightness temperature; however, the relationship between frequency and atmospheric height must be inferred from atmospheric models.

\subsection{Coronal Gyroresonance Emission}\label{sec:gyrores}
The enhanced magnetic field strengths above active regions may cause the corona to become optically thick due to gyroresonant opacity, resulting in bright spots at coronal temperatures of $\sim10^6$\,K on the $10^4$-K stellar disk.  In a given field strength $B$ (in G), free electrons absorb and emit at harmonics of this gyromagnetic frequency:
\begin{equation}
\nu = s \nu_B = s B \left( 2.8 \mathrm{\,MHz} \right) ,
\end{equation}
where the solar corona typically becomes optically thick at low harmonics ($s\sim3-5$).  For gyroresonant radiation in the $s=3$ harmonic to contribute to the observed stellar 34.5-GHz radiation requires the presence of coronal magnetic field strengths of 4.1\,kG; 34.5-GHz emission in the lowest harmonic would require coronal field strengths of 12.3 kG.

Our data constrain the covering fraction $f_{15}$ of coronal bright spots at 15\,GHz, i.e. the covering fraction of $\sim$1.8-kG magnetic fields in the corona. We model the 15-GHz disk-averaged brightness temperature $T_{15}$ using a stellar disk with brightness temperature $T_{chrom}$, which is uniform except where the chromosphere is obscured by coronal bright spots with temperature $T_{cor}$:
\begin{equation}
T_{15} = \left( 1 - f_{15} \right) T_{chrom} + f_{15}~T_{cor} .
\end{equation}
If we assume that the coronal contribution at 34.5\,GHz is negligible, then the measured 34.5-GHz brightness temperature is a direct measurement of $T_{chrom}$.  Combining that with an assumed coronal temperature of $T_{cor}=10^6$\,K, we can convert our upper limits on 15-GHz flux density to upper limits on $f_{15}$.  We obtain 95\%-confidence upper limits on 15-GHz coronal covering fraction of 1.1\% for $\tau$ Cet and 0.8\% for $\eta$ Cas A.

Although our data alone cannot rule out a coronal contribution to the detected 34.5-GHz emission, a significant coronal contribution at 34.5\,GHz is unlikely considering the solar case. \cite{white2004}, and in personal communication \cite{white_private}, report the disk-integrated solar brightness temperature spectrum observed by Nobeyama at a consecutive solar minimum and maximum.  These spectra show no significant variation at 35\,GHz over the solar cycle.  From this we infer that coronal bright spots, likely optically-thin at these high frequencies, contribute only a small, difficult-to-detect fraction of the disk-integrated 35-GHz solar emission.  Since the observed stars have similar levels of magnetic activity to the Sun, coronal emission above active regions likely does not contribute significantly to the stellar 35-GHz emission.  However, if it does, this would provide a diagnostic for tracking stellar cycles, since the active region coverage should vary significantly between stellar minimum and maximum.

Gyroresonant emission from a region with uniform magnetic field can have a high degree of circular polarization, depending on the viewing angle.  The sense of circular polarization is determined by the line-of-sight magnetic field direction.  The gyroresonant contribution to the emission, averaged over the stellar disk, will have lower circular polarization than the emission from individual regions because regions of opposite magnetic polarity cancel each other out.  In a given hemisphere on the Sun, sunspots of one polarity tend to have stronger magnetic fields, which means that the disk-integrated gyroresonant emission from the star may have significant circular polarization when viewed from high latitudes.

The detection of circular polarization in stellar microwave emission would be a smoking gun for a gyroresonant contribution.  Our observations are consistent with zero circular polarization at 34.5\,GHz (refer to Section \ref{section:detections} for constraints on $r_c$), as expected since gyroresonance contributes only a small fraction of the solar 35-GHz emission.  If future, deeper observations of these stars below 10-15\,GHz (the frequency below which gyroresonance contributes significantly to the disk-integrated quiet Sun emission) can detect circular polarization, flips in the sense of this polarization could then be used to track polarity reversals due to the stellar magnetic activity cycle.

\subsection{Coronal Free-Free Emission}

As mentioned above, coronal radiation is unlikely to contribute a significant fraction of the observed 34.5-GHz emission.  In the case of the Sun, the corona is optically-thin to free-free emission above a few GHz.  Optically-thin free-free emission has a nearly flat spectrum, so the rising spectra observed in $\tau$ Cet and $\eta$ Cas A from 15 to 34.5\,GHz imply that the 34.5-GHz emission is not dominated by coronal bremsstrahlung.  One of the sources of coronal bremsstrahlung is the stellar wind, so these observations also place upper limits on the mass loss rates from these stars.

\subsubsection{Stellar Wind Emission}

The diffuse corona that lies on open field lines (i.e., not above active regions) flows into the stellar wind.  This gas contributes a small fraction of the net free-free radio emission calculated from the X-ray emission measure, since most gas in the solar corona is magnetically bound to the star.

We can estimate expected levels of free-free radio emission from the magnetically-open corona by assuming a stellar wind with solar-like properties.
The Sun loses mass through the solar wind at a rate of 2-3$\times10^{-14}$\,M$_\odot$\,yr$^{-1}$.  The solar wind has a fast component and a slow component, as reviewed in \cite{aschwanden2001}.  The base of the solar wind has coronal temperatures, typically around 1\,MK.  For the following calculations, we scale equations to an ``average'' solar wind speed of 500\,km\,s$^{-1}$ and a wind temperature of 1\,MK, and an ionized mass loss rate of $10^{-14}$\,M$_\odot$\,yr$^{-1}$. We assume a spherically-symmetric, constant-velocity wind with a steady mass loss rate, implying a $1/r^2$ density profile.  These simplifying assumptions, combined with the assumed solar wind properties, should provide an order-of-magnitude estimate of the stellar wind radio flux density.

Using Equation 20 of \cite{dulk1985} to calculate the optical depth of a stellar wind with solar-like properties, we obtain:
\begin{align}
\tau = & \left( 2.0 \times 10^{-10} \right) \left( \frac {\nu} {34.5\mathrm{~GHz}} \right) ^ {-2} \left( \frac {T} {10^6 \mathrm{~K}} \right) ^ {-3/2} \ldots\nonumber \\
& \left( \frac {r} {R_\odot} \right) ^ {-3} \left( \frac {\dot{M}_{ion}} {10^{-14} \mathrm{~M_\odot\,yr^{-1}}} \right) ^ 2 \left( \frac {v_w} {500 \mathrm{~km~s^{-1}}} \right) ^ {-2},
\end{align}
where $r$ is the distance of the line of sight from the center of the star and $\dot{M}_{ion}$ is the ionized mass loss rate (since only ionized gas contributes to the free-free emission).  The 34.5-GHz optical depth is extremely low for a solar-like wind.

Since the wind is optically thin, the radio luminosity is obtained by integrating the free-free emissivity over the wind volume (from the stellar surface to infinity), yielding an expected 34.5-GHz flux density of: 
\begin{align}
S_\nu = & \left( 2.9 \times 10^{-5} \mathrm{{\mu}Jy} \right)
        \left( \frac {T} {10^6 \mathrm{~K}} \right) ^ {-1/2}
	\left( \frac {R_*} {R_\odot} \right) ^ {-1} \ldots \nonumber \\
        & \left( \frac {\dot{M}_{ion}} {10^{-14} \mathrm{~M_\odot\,yr^{-1}}} \right) ^ 2
        \left( \frac {v_w} {500 \mathrm{~km~s^{-1}}} \right) ^ {-2}
        \left( \frac {d} {1 \mathrm{~pc}} \right) ^ {-2}
\label{eq:wind_flux}
\end{align} 
for frequencies near 34.5\,GHz and electron temperatures near $10^6$\,K.  More general formulas must adjust for the frequency and temperature dependence of the Gaunt factor (e.g. Equation 24b in \cite{gudel2002}).

Equation \ref{eq:wind_flux} makes it apparent that a stellar wind will not contribute detectably to the stellar radio emission at 34.5\,GHz, unless any of these stars have a mass loss rate more than 1000 times the solar mass loss rate.  \cite{wood2005} report stellar mass loss rates (measured using astrospheric absorption of Ly $\alpha$) of up to 100 times solar for low-mass main sequence stars, but the mass loss rates they measure for stars with solar levels of activity are comparable to the solar mass loss rate.  This stands in contrast to radiation pressure-driven winds from massive stars, which can produce radio emission of 100s of $\mu$Jy or more at kiloparsec distances \citep{scuderi1998}.

\subsection{Gyrosynchrotron Emission}

\cite{linsky1983} proposed that gyrosynchrotron emission dominates the radio luminosity of active stars, in order to explain stellar disk-averaged radio brightness temperatures of greater than $10^8$\,K, signicantly hotter than the coronal temperatures measured in X-rays.  Such high radio brightness temperatures stand in contrast to the $10^4$-K temperatures observed in $\tau$ Cet, $\eta$ Cas A, and 40 Eri A.  We cannot definitively rule out a gyrosynchrotron contribution on the basis of stellar disk-averaged brightness temperature, since the gyrosynchrotron source could be optically thin or cover only small regions of the star, both of which would reduce the disk-averaged brightness temperature, but gyrosynchrotron emission is not required to explain the observed stellar flux densities.  Additionally, the Sun does not flare at a sufficient rate to produce the ``quiescent'' non-thermal gyrosynchrotron microwave radiation observed in magnetically active stars.  Since the observed stars have solar-like levels of magnetic activity and brightness temperatures consistent with thermal emission, gyrosynchrotron radiation is unlikely to contribute appreciably to the detected 35-GHz radio emission.

\section{Conclusions}
\label{section:conclusions}

We have detected 34.5-GHz radio emission from solar-type stars $\tau$ Cet, $\eta$ Cas A, and 40 Eri A, which all have solar-like levels of magnetic activity.  By analogy to the Sun, this emission is most likely thermal chromospheric emission from the entire stellar disk.  Our data cannot constrain the possibility of a coronal contribution from thermal gyroresonance or free-free above active regions, but a large such contribution is unlikely since solar disk-integrated 35-GHz emission is steady throughout the solar cycle, independent of the number of active regions.  Our sensitive upper limits on 15-GHz flux density for $\tau$ Ceti and $\eta$ Cas A indicate that they have a rising spectrum from 15 to 35\,GHz, consistent with optically-thick thermal chromospheric emission.  These observations constitute the first radio detection of thermal stellar emission from low-mass main-sequence stars.

The disk-averaged brightness temperatures at 34.5\,GHz are of order 10,000\,K, higher than the photospheric $\mathrm{T_{eff}}$, providing independent confirmation of the temperature inversion in these stellar atmospheres.  These VLA observations could be combined with ALMA observations to build a brightness temperature spectrum $T_b(\nu)$ for each star.  The brightness temperature spectra can be used to constrain the chromospheric density and temperature profiles predicted by model atmospheres.  These constraints on atmospheric models could be further strengthened by the addition of far-infrared data, which probe the temperature minimum.  Most of our understanding of stellar chromospheres currently comes from UV line emission, which is optically thin across the chromosphere.  Because radio emission becomes optically thick in the chromosphere, with different frequencies becoming opaque at different depths, a radio brightness temperature spectrum directly measures the electron temperature across different chromospheric layers.

Next-generation radio observatories will enable the detection of thermal quiescent emission from solar-type stars at even lower frequencies, probing optically thick regions at coronal temperatures.  As discussed in Section \ref{sec:gyrores}, microwave observations of these stars constrain the covering fraction of magnetic fields of a given strength in the corona, where different frequencies probe different magnetic field strengths.  Detecting the emission below 15\,GHz (the frequency below which gyroresonance starts to contribute an appreciable fraction of the quiescent solar emission) would provide a way to measure magnetic field strength and size of active regions, with rotational modulation of the radio emission potentially providing information on spatial distribution of these active regions.

Radio observations of these stars below 10\,GHz would also provide a diagnostic for tracking stellar magnetic activity cycles, like the solar 10.7-cm flux density $F_{10.7}$, which varies by factors of 2 to 3 over the solar cycle.  Of the observed stars, only 40 Eri A has a detected stellar cycle, seen in variation of calcium-II H \& K lines with a 10.1-year period \citep{baliunas1995}.  Increasing the number of detected stellar activity cycles would provide a test of dynamo theory and explore the long-period end of the relationship between stellar rotation and activity cycle.  The mid-frequency Square Kilometer Array (SKA) will achieve 0.5-$\mu$Jy sensitivity at GHz frequencies in one hour on source\footnote{SKA1 Baseline Design, March 2013. \url{https://www.skatelescope.org/wp-content/uploads/2012/07/SKA-TEL-SKO-DD-001-1_BaselineDesign1.pdf}.}; a few hours per year would be sufficient to monitor stellar activity cycles of low-activity solar-type stars out to about 10\,pc.  The SKA will have high enough sensitivity to potentially detect circular polarization, which can be used to track polarity reversals in stellar magnetic fields.  The enhanced sensitivity of the VLA, combined with the powers of ALMA and eventually the SKA, opens up for exploration a new realm of stellar radio luminosities and emission mechanisms.

\section{Acknowledgements}
The National Radio Astronomy Observatory is a facility of the National Science Foundation operated under cooperative agreement by Associated Universities, Inc.

This material is based upon work supported by the National Science Foundation Graduate Research Fellowship under Grant No.\,DGE-1144469.

This research has made use of the SIMBAD database, operated at CDS, Strasbourg, France.

This research benefited from the web page ``Basic Astronomical Data for the Sun''\footnote{\url{https://sites.google.com/site/mamajeksstarnotes/basic-astronomical-data-for-the-sun}} created by Eric Mamajek of the University of Rochester.

J.V. thanks Stephen White and Jeffrey Linsky for giving helpful feedback on the paper and additionally thanks Stephen White for information on the quiescent solar microwave brightness temperature spectrum.

\bibliography{solartype}
\bibliographystyle{apj}

\end{document}